\documentclass[12pt]{article}

\title{Consistency of the Maximum Likelihood Estimator of Evolutionary Tree}

\author{Arindam RoyChoudhury$^{1\ast}$\\
\textit{$^{1}$Department of Biostatistics}\\
\textit{Columbia University, New York NY 10032, USA}\\
$^{\ast}$ corresponding author
}
\date{}

\begin{document}

\bibliographystyle{plain}

\raggedright

\setlength\parindent{0.5in}

\maketitle

\noindent \hspace{0.5 in} Maximum likelihood estimation (MLE) methods are widely used for evolutionary tree. As evolutionary tree is not a smooth parameter, the consistency of its MLE has been a topic of debate. It has been noted without proof that the classical proof of consistency by Wald holds for the MLE of evolutionary tree. Other proofs of consistency under various models were also proposed. Here we will discuss some shortcomings in some of these proofs and comment on the applicability of Wald's proof.

\noindent Key words: consistency, maximum likelihood, evolutionary tree, phylogenetic tree, species tree

\begin{centering}

\large{E}\small{XISTING} \large{P}\small{ROOFS IN THE} \large{L}\small{ITERATURE}

\end{centering}

\vspace{0.2 in}

\begin{centering}

\textit{The Proof by \cite{y94,r97,f04}}

\end{centering}

\vspace{0.1 in}

\noindent \hspace{0.5 in} A proof was outlined by \cite{y94}. Then it was formalized by \cite{r97} and later restated in a simpler form by \cite{f04}. We will use the form provided by \cite{f04} to point out the subtle mathematical deficiency in the proof. The other versions of the proof suffer from the same deficiency. 

\noindent \hspace{0.5 in} The proof provided by \cite{f04} (pages 271-272) is as follows. Suppose that there are $m$ possible character patterns $x_1,x_2,\dots,x_m$. The data consist of $n$ independently observed patterns where $x_i$ occurs $n_i$ times. ($\sum_{i=1}^m n_i = n$.)
Then the likelihood of tree ($T$) is
$$ L(T) = \prod_{i=1}^m \mbox{Pr}(x_i ; T)^{n_i}, $$
where $\mbox{Pr}(x_i ; T)$ is the probability of observing $x_i$ under the tree-parameter value $T$.
Thus,
\begin{eqnarray}
\frac{1}{n}\log L(T) & = & \sum_{i=1}^m f_i^{\ast} \log p_i, \label{eq:f04_log_like}
\end{eqnarray}
where $f_i^{\ast} = n_i/n$ and $p_i = \mbox{Pr}(x_i ; T)$.

\noindent \hspace{0.5 in} Suppose that the true parameter value is $T_0$ 
and the probability of observing the $i$-th pattern is $p_i(0)$ under $T_0$. Then for another set of
probability $p_i(j)$ (under another parameter value $T_j$):
\begin{eqnarray}
\sum_{i=1}^m p_i(0) \log p_i(0) > \sum_{i=1}^m p_i(0) \log p_i(j)
\nonumber
\end{eqnarray}
(from Gibbs' inequality).

\noindent \hspace{0.5 in} After this point the proof is supposed to proceed as follows,
$$ \mbox{E} \Biggl( \frac{1}{n} \log \Bigl(L(T_0)\Bigr) \Biggr) - \mbox{E} \Biggl( \frac{1}{n} \log \Bigl(L(T_j)\Bigr) \Biggr) = \sum_{i=1}^m p_i(0) \log p_i(0) - \sum_{i=1}^m p_i(0) \log p_i(j).$$
Therefore,
$$ \frac{1}{n} \log \Bigl(L(T_0)\Bigr) - \frac{1}{n} \log \Bigl(L(T_j)\Bigr) \to_{\mbox{\scriptsize{a.s.}}} \sum_{i=1}^m p_i(0) \log p_i(0) - \sum_{i=1}^m p_i(0) \log p_i(j) > 0$$
(using the Strong Law of Large Numbers.)
Thus, there exist $N_j$, such that for all $n \geq N_j$, 
\begin{eqnarray}
L(T_0) > L(T_j)
\label{eq:obs_inq}
\end{eqnarray}
with probability 1. 

\noindent \hspace{0.5 in} Now, if the parameter space was finite, then this would suffice. We could number the rest (other than $T_0$) of the parameters as $T_1,\dots,T_p$.
Then for each $1 \leq j \leq p$, we could find $N_j$ such that Eq. (\ref{eq:obs_inq}) holds. If we take
$$N \geq \max_{1 \leq j \leq p} N_j,$$
then for all $n \geq N$, $L(T_0)$ is the maximum likelihood with probability 1 (meaning that the MLE $\hat{T} = T_0$ with probability 1, and hence the MLE is consistent.)
However, the parameter space is infinite, as there are infinitely many possible values for each branch length.
To use a similar argument for infinite parameter space, we have to consider 
\begin{eqnarray}
N' \geq \sup_{x \in A} N_x,
\label{eq:sup}
\end{eqnarray}
where $A$ is an indexing of the parameter space (excluding $T_0$).
But, $\sup_{x \in A} N_x$ may be $\infty$. Thus, our previous argument will not work. 

\cite{f04} stated that some other conditions are needed to ensure the actual convergence of MLE to the true tree. He, however, did not provide those conditions. The intuitive argument of \cite{y94} does not address the mathematical problem described above.

\vspace{0.2 in}

\begin{centering}

\textit{The Proof by \cite{r01}}

\end{centering}

\vspace{0.1 in}

\noindent \hspace{0.5 in} \cite{r01} numbered the tree-topologies as $1,2,\dots,K$. The proof by \cite{r01} treats the numbered topologies and other numeric parameters as points in Cartesian space, and attempts to establish that the parameter space thus constructed is closed. It argues that as each parameter $\theta'$ (including the numeric tree-topology parameter) can be transformed so that $ 0 \leq \theta' \leq c$ for a constant $c$, the transformed parameter space is closed. However, this argument is not enough to prove that a set is closed. \cite{r01} considered only completely bifurcating trees in the parameter space. Thus for a given internal branch the set of all possible branch lengths does not include 0. Therefore, the parameter space is not closed.

\noindent \hspace{0.5 in} Including all trees in the parameter space may appear to be an easy remedy to this. But, that leads to some inherent difficulties, as described below.

\noindent \hspace{0.5 in} Any multifurcated tree-topology $Y$ can be obtained by collapsing one or more internal branches of at least two distinct completely bifurcating tree-topologies $Y_1$ and $Y_2$. Therefore, one can have two sequence of trees $S_1$ and $S_2$, having tree-topologies $Y_1$ and $Y_2$ respectively, such that they converge to a common limiting tree with tree-topology $Y$. Therefore, to be closed, $Y$ has to have same numeric value as both $Y_1$ and $Y_2$, which is impossible under the numbering scheme of \cite{r01}. 

An identifiability problem with the proof was also discussed by \cite{aea08}. (Identifiability is a requirement for consistency as unidentifiable parameter space could lead to serious estimation problems; see, for example, \cite{ms07}.)

\vspace{0.2 in}

\begin{centering}

\textit{Applicability of the proof by \cite{w49}} 

\end{centering}

\vspace{0.1 in}

\noindent \hspace{0.5 in} As mentioned before, \cite{f73} noted that the proof by \cite{w49} can be used to proof the 
consistency. \cite{sea01} debunked the claim of some works (such as \cite{y96} and \cite{f99}) that argued that the differentiability and continuity of likelihood (as a function of the parameters) is a requirement in the proof by \cite{w49}. However, Assumption 7 (the parameter space has to be a closed subset of a Cartesian space) of \cite{w49} appears not to hold. (For a detailed analysis of the assumptions of Wald, 1949 in the context of genetic data, see Rogers, 2001).

\noindent \hspace{0.5 in} \cite{w49} noted that Assumption 7 is unnecessarily binding and stated an alternative Assumption 9 in \cite{w49}. The alternative
involves a condition that a metric has to be defined in the parameter 
space, under which all closed and bounded sets are compact (Assumption 9(i,iv)). (The other conditions in Assumption 9 will be satisfied by any reasonable model.)
 
\noindent \hspace{0.5 in} There are two kinds of parameters associated with a tree: branch-specific (such as branch length) and non-branch-specific (the overall parameters that are not associated with a particular branch). The set of all possible values of the branch-specific parameters can be shown to have one-to-one correspondence with a complete real metric space, where the natural metric conforms to the natural concept of distance in evolutionary trees (\cite{bea01}). (In a real valued complete metric space all closed and bounded sets are compact; moreover, the definition of closed set would be preserved from the tree-space to the metric space as the concept of distance is preserved.) Therefore, if all the non-branch-specific parameters also form a real complete metric space, then the MLE is consistent for evolutionary tree.

\vspace{0.3 in}

\begin{centering}

\large{D}\small{ISCUSSION}

\end{centering}

\vspace{0.1 in}

\noindent \hspace{0.5 in}  We have pointed out mathematical shortcomings in some of the existing proofs of consistency of MLE of evolutionary tree. We have also established that Wald's proof of consistency \cite{w49} can indeed be used to prove the consistency of MLE of evolutionary tree as suggested by \cite{f73}.

\noindent \hspace{0.5 in} A simulation-based verification of consistency is not feasible for trees with a large number of taxa. As trees with a larger number of taxa can potentially introduce structures that are not present in fewer-taxa trees, a simulation-based verification of consistency in the latter is not enough to assert the consistency in the former. Therefore, a theoretical proof of consistency (such as the one noted in this work) is required to ensure that the MLE of evolutionary tree with any number of taxa is indeed consistent.

\vspace{0.3 in}

\begin{centering}

\large{A}\small{CKNOWLEDGMENTS}

\end{centering}

\vspace{0.1 in}

\noindent \hspace{0.5 in} The author is grateful to Joseph Felsenstein, John Wakeley and Adam Seipel for their valuable suggestions.  

\renewcommand*{\refname}{}

\vspace{0.3 in}

\begin{centering}

\large{R}\small{EFERENCES}

\end{centering}

\vspace{-0.9 in}

\bibliography{arindam}

\end{document}